\documentclass[12pt,preprint,notoc,nohyper]{CKVPlane2} 

\usepackage{epsfig,multicol,bbm}

\newcommand\fverb{\setbox\pippobox=\hbox\bgroup\verb}
\newcommand\fverbdo{\egroup\medskip\noindent%
            \fbox{\unhbox\pippobox}\ }
\newcommand\fverbit{\egroup\item[\fbox{\unhbox\pippobox}]}
\newbox\pippobox

\title{Conformal motions in plane
symmetric static spacetimes}

\author{K. Saifullah$^a$\footnote{\emph{On leave
from:} Centre for Advanced Mathematics and Physics, National
University of Sciences and Technology, Rawalpindi, Pakistan,
\emph{and} Department of Mathematics, Quaid-i-Azam University,
Islamabad, Pakistan. (Electronic address: saifullah@qau.edu.pk)}
~ and Shair-e-Yazdan$^b$ \\

$^a$School of Mathematical Sciences, Queen Mary, University of
London, London, United Kingdom \\
$^b$S-2/95, Saudabad, Malir, Karachi, Pakistan \\}

\preprint{}  

\abstract{In this paper, conformal motions are studied in plane
symmetric static spacetimes. The general solution of conformal
Killing equations and the general form of the conformal Killing
vector for these spacetimes are presented. All possibilities for the
existence of conformal motions in these spacetimes are exhausted.}

\dedicated{Dec 2007}

\begin{document}

\section{Introduction and Motivation}

In addition to isometries there are other types of motions also
which are very useful as far as the four dimensional Lorentzian
metrics, their properties and their applications to mathematical
physics are concerned. Conformal motions or conformal Killing
vectors (CKVs) are motions along which the metric tensor of a
spacetime remains invariant upto a scale factor. A conformal vector
field can be defined\cite{2r,9r} as a global smooth vector field
$\mathbf{\xi}$ on a manifold, $M$, such that for the metric $g_{ab}$
in any coordinate system on $M$
\begin{equation}
\xi_{a;b}=\phi g_{ab}+F_{ab} ,  \label{s38}
\end{equation}%
where $\phi :M\longrightarrow R$ is the smooth conformal function of
$\mathbf{\xi,}$ $F_{ab}\left( =-F_{ba}\right)$ is the conformal
bivector of $\mathbf{\xi}$. This is equivalent to
\begin{equation}
\pounds _{\mathbf{\xi}}g_{ab}=2g_{ab}\phi(t,x,y,z) , \label{s36}
\end{equation}
where $\pounds _{\mathbf{\xi}}$ represents the Lie derivative with
respect to $\mathbf{\xi}$.

In explicit form we can write the above equation as,
\begin{equation}
g_{ab,c}\xi^{c}+g_{cb}\xi^{c},_{a}+g_{ac}\xi^{c},_{b}=2g_{ab}\phi
(t,x,y,z) .  \label{s37}
\end{equation}

The \textquotedblleft , \textquotedblright  represents the partial
derivative with respect to coordinates $x^{a}(x^{0}=t, x^{1}=x,
x^{2}=y, x^{3}=z)$. Here, if $\phi$ is constant, $\xi$ are called
homothetic motions or homothetic  vector (HV) fields, and if it is
zero we get Killing vector (KV) fields. The KVs, HVs and CKVs form
finite dimensional Lie algebras. For a basis
$\{\mathbf{X}_{i},i=1,...,n\}$ of a Lie algebra, we can always write
the Lie bracket

\begin{equation}
\left[ \mathbf{X}_{k},\mathbf{X}_{l}\right]
=C_{kl}^{j}\mathbf{X}_{j}
\,\,\,\,\,\,\,\,\,\,\,\,\,\,\,\,\,\,\,\,\,\,\,\,\,\,\,\,\,%
\,C_{kl}^{j}=-C_{lk}^{j},
\end{equation}
which is bilinear, antisymmetric and satisfies the Jacobi identity.
Here $C_{kl}^{j}$ are the structure constants which completely
characterize the Lie algebra.

Here we state some of the well-known results\cite{2r,9r} regarding
dimensionality of the Lie algebras for motions.

\textbf{Theorem 1}:

\textit{The Lie algebras of KVs and HVs are finite dimensional. For
an n-dimensional manifold M admitting a metric of any signature, the
dimension of the algebra of KVs is $\leq\frac{n(n+1)}{2}$ and that
of HVs is $\leq\frac{n(n+1)}{2}+1$.}

\pagebreak

\textbf{Theorem 2}:

\textit{The set of conformal vector fields on a spacetime is
finite-dimensional and its dimension is $\leq15$. If this maximum
number is attained, the spacetime is conformally flat. If it is not
conformally flat then the dimension of the set of CKVs is $\leq7$.}

Conformal motions are determined by the arbitrary constants
appearing in the vector field $\mathbf{\xi }=\xi ^{a}\partial
/\partial x^{a}$ when $\phi=\phi (t,x,y,z)$. The study of the
symmetry groups of a spacetime is a useful tool in constructing
spacetime solutions of EFEs and also classifying the known solutions
according to their Lie algebras, or structures generated by these
symmetries. They have physical significance as they generate motion
along null geodesics for massless particles. CKVs have been studied
for various classes of spacetimes including Minkowski\cite{5r},
Friedmann-Robertson-Walker\cite{6r} and pp-waves\cite{7r}. The
general solution and classification of conformal motions in static
spherical spacetimes has also been carried out\cite{11r}. While much
work has been done on isometries and homotheties, comparatively
little is known about conformal symmetries. This is at least in
part, due to the difficulty in solving the conformal Killing
equations which contain conformal factor $\phi $, which is in
general a function of all coordinates\cite{8r}.

In this paper we study the conformal motions or CKVs of plane
symmetric static spacetimes. We first find the general solution of
the equations of conformal motions in these spacetimes, together
with the general form of the CKV and the conformal factor. Then we
provide the complete classification of CKVs.

In Cartesian coordinates the general form of static plane symmetric
spacetime is
\begin{equation}
ds^{2}=e^{2\nu(x) }dt^{2}-dx^{2}-e^{2\mu(x) }\left(
dy^{2}+dz^{2}\right) \label{a29}
\end{equation}
The general form of KVs, $\mathbf{K}$, for this metric is given
by\cite{3r}
\begin{eqnarray}
K^{0} &=&e^{2\left( \mu -\nu \right) }\left\{ \frac{1}{2}\
\dot{A}_{1}\left( t,x\right) \left( z^{2}+y^{2}\right)
+\dot{A}_{2}\left( t,x\right) z+\dot{A}
_{3}\left( t,x\right) y\right\} +K\left( t,x\right) ,  \label{a30} \\
K^{1} &=&-e^{2\mu }\left\{ \frac{1}{2}A_{1}^{\prime }\left(
t,x\right) \left( z^{2}+y^{2}\right) +A_{2}^{\prime }\left(
t,x\right) z+A_{3}^{\prime
}\left( t,x\right) y\right\} +L\left( t,x\right) ,  \label{a31} \\
K^{2} &=&+\frac{c_{3}}{2}\left( z^{2}-y^{2}\right) +yzc_{2}+c_{1}z+\
\ \
A_{1}\left( t,x\right) y+A_{3}\left( t,x\right) ,  \label{a32} \\
K^{3} &=&+\frac{c_{2}}{2}\left( z^{2}-y^{2}\right)
-yzc_{3}-c_{1}y+A_{1}\left( t,x\right) z+A_{2}\left( t,x\right) ,
\label{a33}
\end{eqnarray}
where $A_{i}\left( t,x\right)$, $K\left( t,x\right)$ and $L\left(
t,x\right)$ satisfy some differential constraints.

This classification reproduces the well-known static plane symmetric
solutions, giving anti-de Sitter metric in the form
\begin{equation}
ds^{2}=e^{\frac{2x}{x_0}}\left( dt^{2}-dz^{2}-dy^{2}\right) -dx^{2}
. \nonumber
\end{equation}
It also reprodues the 5-dimensional isometries

\begin{eqnarray}
K^{0} &=&x_0 \nu ^{\prime} c_{5}+c_{1}, \nonumber \\
K^{1} &=&-c_{5}{x_0}, \nonumber \\
K^{2} &=&c_{5}y+c_{4}z+c_{2}, \nonumber \\
K^{3} &=&c_{5}z-c_{4}y+c_{3}, \nonumber
\end{eqnarray}
admitted by the class of metrics

\begin{equation}
ds^{2}=e^{\nu }dt^{2}-dx^{2}-e^{\frac{x}{x_0}}\left(
dz^{2}+dy^{2}\right) , \nu ^{\prime \prime }\neq 0 . \nonumber
\end{equation}

There were three new metrics with $6$ isometries. These were planar
analogues of the Bertotti-Robinson metrics and two other similar
metrics\cite{16r}. They have $\mu (x)$ $=0$ and
\begin{eqnarray}
\nu \left( x\right) &=&\ln \cosh ^{2}\alpha x, \nonumber \\
\nu \left( x\right) &=&e^{2\alpha x}, \nonumber \\
\nu \left( x\right) &=&\ln \cos ^{2}\alpha x, \nonumber
\end{eqnarray}
and the corresponding KVs are

\begin{eqnarray}
K&=&\left[ c_{0}-\tanh \alpha x\left( c_{4}\sin \alpha t-c_{5}\cos
\alpha t\right) \right] \partial /\partial t+\left( c_{4}\sin \alpha
t+c_{5}\cos \alpha t\right) \partial /\partial x \nonumber \\
&&+ \left( c_{1}+c_{3}z\right) \partial /\partial y+\left(
c_{2}-c_{3}y\right)
\partial /\partial z, \nonumber \\
K&=&\left[ c_{0}-\alpha \left( c_{4}t^{2}+c_{5}t\right) \right]
\partial /\partial t+\left( 2c_{4}t+c_{5}\right) \partial /\partial
x+\left(c_{1}+c_{3}z\right) \partial /\partial y \nonumber \\
&&+ \left(
c_{2}-c_{3}y\right) \partial /\partial z, \nonumber \\
K&=& \left[c_{0}-\left(c_{4}\sin \alpha t+c_{5}\cos \alpha t\right)
\tan \alpha x\right]\partial /\partial t+\left( c_{4}\cos \alpha
t+c_{5}\sin \alpha t\right) \partial /\partial x\nonumber \\
&&+ \left( c_{1}+c_{3}z\right)
\partial /\partial y+\left( c_{2}-c_{3}y\right)
\partial /\partial z, \nonumber
\end{eqnarray}
where $c_{i}$ are arbitrary constants.

\section{General Form of Conformal Motions}

The conformal Killing Eqs. (\ref{s37}) for the metric given by
(\ref{a29}) represent a system of ten coupled non homogeneous first
order partial differential equations. As the metric is diagonal, we
note that from Eqs. (\ref{s37}) for $a=b$ the only non-zero
contributions from the second and the third term on the left hand
side will come when $c=a=b$. When $a\neq b$ these terms will
contribute only when $c=b$ and $c=a$, respectively. Thus by dropping
the summation convention these can be written as

\begin{eqnarray}
g_{aa}^{\prime} \xi^{1}+2g_{aa}\xi_{,a}^{a} =2g_{aa}\phi (t,x,y,z) ,
\ \ \ \ \ (a=0,1,2,3) \label{c1} \\
g_{bb} \xi_{,a}^{b}+g_{aa}\xi_{,b}^{a} =0 , \ \ \ \ \ (a,b=0,1,2,3;
a \neq b) . \label{c2}
\end{eqnarray}
Here Eqs. (\ref{c1}) give four equations and Eqs. (\ref{c2}) are six
equations. These are to be solved\cite{thesis} to give six unknown
functions, $\nu (x)$, $\mu (x)$ and $\xi^{\alpha }=\xi
^{\alpha}(x^{a})$ and the conformal factor $\phi(x^{i})$. We first
write Eqs. (\ref{c2}) for $a=2, b=0$ and $a=3, b=0$ and
differentiate them with respect to $z$ and $y$ respectively to
obtain

\begin{equation}
e^{2\nu }\xi _{,23}^{0}-e^{2\mu }\xi _{,03}^{2}=0,  \label{t11}
\end{equation}
\begin{equation}
e^{2\nu }\xi _{,23}^{0}-e^{2\mu }\xi _{,02}^{3}=0.  \label{t12}
\end{equation}
Now we write Eq. (\ref{c2}) for $a=2, b=3$ and differentiate with
respect to $t$ to get
\begin{equation}
\xi_{,03}^{2}=-\xi_{,02}^{3}.  \label{t12a}
\end{equation}
Using Eq. (\ref{t12a}) \ in Eq. (\ref{t12}), we get
\begin{equation}
e^{2\nu }\xi _{,23}^{0}+e^{2\mu }\xi _{,03}^{2}=0.  \label{t12b}
\end{equation}
Adding Eqs. (\ref{t11}) and (\ref{t12b}) we get
\begin{equation}
\xi _{,23}^{0}=0.  \label{t13}
\end{equation}
Similarly, from Eqs. (\ref{c2}) for $a=1, b=2$ and $a=1, b=3$ we get
\begin{equation}
\xi _{,23}^{1}=0.  \label{t16}
\end{equation}
Comparing Eqs. (\ref{c1}) for $a=1$ and $a=2$ gives
\begin{equation}
\xi _{,3}^{2}=B_{1}\left( t,x,z\right) y+B_{2}\left( t,x,z\right) ,
\label{t20}
\end{equation}
where $B_{1}$ and $B_{2}$ are functions of integration. Using this
in Eqs. (\ref{c2}) for $a=2, b=3$ and integrating with respect to
$y$ yields
\begin{equation}
\xi ^{3}=-\frac{1}{2}B_{1}\left( t,x,z\right) y^{2}-B_{2}\left(
t,x,z\right) y+B_{3}\left( t,x,z\right).  \label{t22}
\end{equation}
Substituting from Eqs. (\ref{t20}) and (\ref{t22}) in Eqs.
(\ref{c1}) for $a=2$ and $a=3$, respectively, comparing the
coefficients of $y^{2}$, $y$ and the terms independent of $y$, and
integrating with respect to $z$ yields
\begin{eqnarray}
B_{1}\left( t,x,z,\right) &=&F_{1}\left( t,x\right) z+F_{2}\left(
t,x\right)
,  \label{t28} \\
B_{2}\left( t,x,z,\right) &=&F_{3}\left( t,x\right) z+F_{4}\left(
t,x\right)
,  \label{t29} \\
B_{3}\left( t,x,z\right) &=&F_{1}\left( t,x\right) \frac{z^{3}}{6}%
+F_{2}\left( t,x\right) \frac{z^{2}}{2}+A_{1}\left( t,x\right)
+A_{2}\left( t,x\right) ,  \label{t30}
\end{eqnarray}%
where $F_{i}(t,x)$ and $A_{i}(t,x)$ are functions of integration.
Using these values of $B_{1},B_{2}$ and $B_{3}$ we obtain from Eqs.
(\ref{t20})- (\ref{t22})

\begin{equation}
\xi ^{2}=\left\{ \frac{1}{2}F_{1}\left( t,x\right) z^{2}+F_{2}\left(
t,x\right) z\right\} y+\frac{1}{2}F_{3}\left( t,x\right)
z^{2}+F_{4}\left( t,x\right) +B_{4}\left( t,x,y\right) , \label{t34}
\end{equation}
and
\begin{eqnarray}
&\xi ^{3} =-\frac{1}{2}\left\{ F_{1}\left( t,x\right) z+F_{2}\left(
t,x\right) \right\} y^{2}-\left\{ F_{3}\left( t,x\right)
z+F_{4}\left(
t,x\right) \right\} y  \nonumber \\
&+F_{1}\left( t,x\right) \frac{z^{3}}{6}+F_{2}\left( t,x\right)
\frac{z^{2}}{2}+A_{1}\left( t,x\right) +A_{2}\left( t,x\right) .
\label{t33}
\end{eqnarray}
where $B_{4}$ is a function of integration. Substituting these
values of $\xi ^{2}$ and $\xi ^{3}$ in Eqs. (\ref{c1}) for $a=2$ and
$a=3$ and comparing yields

\begin{equation}
B_{4}\left( t,x,y\right) =-\frac{1}{6}F_{1}\left( t,x\right) y^{3}-\frac{1}{2%
}F_{3}\left( t,x\right) y^{2}+A_{1}\left( t,x\right) y+A_{3}\left(
t,x\right) ,  \label{t36}
\end{equation}
where $A_{3}\left( t,x\right) $ is a function of integration.
Substituting $\xi ^{3}$ in Eqs. (\ref{c2}) for $a=0, b=3$ and $a=1,
b=3$ respectively and integrating the resulting equations yields
\begin{eqnarray}
\xi ^{0}  = e^{2( \mu -\nu ) }[-\{ \frac{1}{4}\dot{F}_{1}( t,x)
z^{2}+\frac{1}{2}\dot{F} _{2}( t,x) z\} y^{2}-\{
\frac{1}{2}\dot{F}_{3}( t,x) z^{2}+\dot{F}_{4}(
t,x) z\} y  \nonumber \\
 + \frac{1}{24}\dot{F}_{1}( t,x) z^{4}+\frac{1}{6}\dot{F}_{2}( t,x)
z^{2}-\dot{A}_{2}( t,x)  + \frac{1}{2} \dot{A}_{1}( t,x)
y^{2}+\dot{A}_{3}( t,x) y] +A_{0}( t,x) \label{t38}
\end{eqnarray}

\begin{eqnarray}
\xi ^{1}=-e^{2\mu }[-\{\frac{1}{4}F_{1}^{\prime }( t,x)
z^{2}+\frac{1}{2}F_{2}^{\prime }( t,x) z\} y^{2}-\{ \frac{1}{2}%
F_{3}^{\prime }( t,x) z^{2}+F_{4}^{\prime }(
t,x) z\} y  \nonumber \\
+\frac{1}{24}F_{1}^{\prime }( t,x) z^{4}+\frac{1}{6}F_{2}^{\prime }(
t,x) z^{2}-A_{2}^{\prime }( t,x) + \frac{1}{2}A_{1}^{\prime }( t,x)
y^{2}+A_{3}^{\prime }( t,x) y] +A_{4}( t,x) \label{t39}
\end{eqnarray}
where $A_{0}(t,x)$ and $A_{4}(t,x)$ are functions of integration.

Substituting these values of $\xi^i$ in Eqs. (\ref{c1}) and
(\ref{c2}) for checking consistency yields

\begin{equation}
\dot{F}_{i}\left( t,x\right) =0=F_{i}^{\prime }\left( t,x\right) ,\
(i=1,2,3,4). \nonumber
\end{equation}

Thus
\begin{equation}
F_{1}\left( t,x\right) =c_{0}, F_{2}\left( t,x\right) =c_{2},
F_{3}\left( t,x\right) =c_{3}, F_{4}\left( t,x\right) =c_{4},
\nonumber
\end{equation}
where $c_{i}$, are arbitrary constants. Hence we obtain the general
form of the CKV as

\begin{eqnarray*}
\mathbf{\xi } &=&\left[ e^{2\left( \mu -\nu \right) }\left\{
\frac{1}{2}\ \dot{A}_{1}\left( t,x\right) \left( z^{2}+y^{2}\right)
+\dot{A}_{2}\left( t,x\right) z+\dot{A}_{3}\left( t,x\right)
y\right\} +A_{0}\left( t,x\right)
\right] \partial /\partial t \\
&&-\left[ e^{2\mu }\left\{ \frac{1}{2}A_{1}^{\prime }\left(
t,x\right) \left( z^{2}+y^{2}\right) +A_{2}^{\prime }\left(
t,x\right) z+A_{3}^{\prime }\left( t,x\right) y\right\} -A_{4}\left(
t,x\right) \right] \partial
/\partial x \\
&&+\left[ \frac{c_{3}}{2}\left( z^{2}-y^{2}\right) +c_{2}yz+c_{4}z+
A_{1}\left( t,x\right) y+A_{3}\left( t,x\right) \right]
\partial
/\partial y\\
&&+\left[ \frac{c_{2}}{2}\left( z^{2}-y^{2}\right)
-c_{3}yz-c_{4}y+A_{1}\left( t,x\right) z+A_{2}\left( t,x\right)
\right]
\partial /\partial z .
\end{eqnarray*}

In order to obtain the explicit form of $\xi ^{a}$ from the above we
need to know the arbitrary functions $A_{i}(t,x),i=0, \dots ,4$. For
this we substitute these equations in the CKV equations (\ref{c1})
and (\ref{c2}) and obtain the following differential
constraints\cite{thesis}.

\begin{eqnarray}
A_{1}^{\prime \prime }\left( t,x\right) +\mu ^{\prime }A_{1}^{\prime
}\left(
t,x\right) &=&0,  \label{t65a} \\
A_{2}^{\prime \prime }\left( t,x\right) +\mu ^{\prime }A_{2}^{\prime
}\left(
t,x\right) &=&-c_{2}e^{-2\mu },  \label{t65b} \\
A_{3}^{\prime \prime }\left( t,x\right) +\mu ^{\prime }A_{3}^{\prime
}\left(
t,x\right) &=&c_{3}e^{-2\mu },  \label{t65c} \\
A_{k}^{\prime }\left( t,x\right) \left[ 2\mu ^{\prime }-\nu ^{\prime
}\right] +e^{-2\nu }\ddot{A}_{k}\left( t,x\right) +A_{k}^{\prime
\prime }\left(
t,x\right) &=&0,\ \ \ \ \ k=1,2,3  \label{t65} \\
\dot{A}_{k}^{\prime }\left( t,x\right) +\left[ \mu ^{\prime }-\nu
^{\prime }\right] \dot{A}_{k}\left( t,x\right) &=&0,\ \ \
k=1,2,3  \label{t66} \\
\mu ^{\prime }A_{4}\left( t,x\right) +A_{1}\left( t,x\right)
-A_{4}^{\prime }\left( t,x\right) &=&0,  \label{t67} \\
e^{2\nu }A_{0}^{\prime }\left( t,x\right) -\dot{A}_{4}\left(
t,x\right) &=&0 ,  \label{t68} \\
\nu ^{\prime }A_{4}\left( t,x\right) +\dot{A}_{0}\left( t,x\right)
-A_{4}^{\prime }\left( t,x\right) &=&0.  \label{t69}
\end{eqnarray}

We also find that the general form of the conformal factor is
\begin{eqnarray}
\phi = -e^{2\mu }\left[ \frac{1}{2}A_{1}^{\prime \prime }\left(
t,x\right) \left( z^{2}+y^{2}\right) +A_{2}^{\prime \prime }\left(
t,x\right) z+A_{3}^{\prime \prime }\left( t,x\right) y\right]
\nonumber \\ - 2\mu ^{\prime }e^{2\mu }\left[
\frac{1}{2}A_{1}^{\prime }\left( t,x\right) \left(
z^{2}+y^{2}\right) +A_{2}^{\prime }\left( t,x\right) z+A_{3}^{\prime
}\left( t,x\right) y\right] +A_{4}^{\prime }\left( t,x\right).
\label{t59a}
\end{eqnarray}

\section{Classification of Conformal Motions}

The problem of finding CKVs in plane symmteric static spacetimes is
now reduced to solving the twelve coupled non linear non homogeneous
second order partial differential Eqs. (\ref{t65a})-(\ref{t69}) to
give seven unknown functions $\nu$, $\mu$ and $A_{i}(t,x), i=0,
...,4$. The arbitrary constants in $\mathbf{\xi}$ will determine the
number of generators of the Lie algebra. We divide our
classification scheme into different cases depending upon whether
one, both or none of the metric coefficients are
constants\cite{thesis}.

Let us first consider the simplest case when both $\mu$ and $\nu$
are constant. We note that the Weyl tensor becomes zero and thus
this is a class of conformally flat spacetimes. The final form of
the CKVs is

\begin{eqnarray*}
\xi ^{0}&=&\left[ \frac{1}{2}c_{5}\left( z^{2}+y^{2}\right) +\left(
c_{2}t+c_{7}\right) z+\left( -c_{3}t+c_{11}\right) y\right]
+c_{1}xt+c_{1}\left( \frac{x^{2}}{2}+ \frac{t^{2}}{2}\right)
+ \\
&& c_{6}t+c_{13}x+c_{15}, \\
\xi ^{1} &=&-\left[ \frac{1}{2}c_{1}\left( z^{2}+y^{2}\right)
+\left( -c_{2}x+c_{9}\right) z+\left( c_{3}x+c_{10}\right) y\right]
+c_{5}tx+c_{1}\left( \frac{x^{2}}{2}+\frac{t^{2}}{2}\right)
+\\
&&c_{6}x+c_{13}t+c_{14}, \\
\xi ^{2} &=&\frac{c_{3}}{2}\left( z^{2}-y^{2}\right)
+c_{2}yz+c_{4}z+\left( c_{1}x+c_{5}t+c_{6}\right)
y+c_{3}\frac{x^{2}}{2}+c_{10}x- \\
&&c_{3}\frac{t^{2}}{2}+c_{11}t+c_{12}, \\
\xi ^{3} &=&\frac{c_2}{2}\left( z^{2}-y^{2}\right)
-c_{3}yz-c_{4}y+\left( c_{1}x+c_{5}t+c_{6}\right) z+c_{2}\left(
\frac{t^{2}}{2}-\frac{x^{2}}{2} \right) + \\
&&c_{7}x+c_{7}t+c_{8},
\end{eqnarray*}
which is the well-known\cite{5r} 15-dimensional Lie algebra of
Minkowski spacetime. The coformal factor from Eq. (\ref{t59a}) takes
the following form in this case

\begin{equation}
\phi \left( t,x,y,z\right) =c_{5}t+c_{1}x-c_{3}y+c_{2}z+c_{6} .
\nonumber
\end{equation}

Next we take $\mu ^{\prime }\neq 0$, $\nu ^{\prime }=0$, and for
simplicity we take $\nu =0$. It is worth noting here that the metric
in this case is conformally related to the one where both $\mu
^{\prime }$ and $\nu ^{\prime}$ are non-zero as we can redefine the
coordinate $x$ to express the metric in former way. The conformal
algebra in the two cases are identical. However, when $\mu ^{\prime
}=\nu ^{\prime}\neq0$, the spacetime is conformally related to
Minkowski spacetime and therefore admits a 15-dimensional conformal
algebra.

Substituting $\nu=0 $ in Eqs. (\ref{t65a})-(\ref{t69}), we have
\begin{eqnarray}
A_{1}^{\prime \prime }\left( t,x\right) +\mu ^{\prime }A_{1}^{\prime
}\left(
t,x\right) &=&0,  \label{t98} \\
A_{2}^{\prime \prime }\left( t,x\right) +\mu ^{\prime }A_{2}^{\prime
}\left(
t,x\right) &=&-c_{2}e^{-2\mu },  \label{t99} \\
A_{3}^{\prime \prime }\left( t,x\right) +\mu ^{\prime }A_{3}^{\prime
}\left(
t,x\right) &=&c_{3}e^{-2\mu },  \label{t100} \\
\dot{A}_{k}^{\prime }\left( t,x\right) +\mu ^{\prime
}\dot{A}_{k}\left( t,x\right) &=&0,\ k=1,2,3  \label{t101} \\
2\mu ^{\prime }A_{k}^{\prime }\left( t,x\right) +\ddot{A}_{k}\left(
t,x\right) +A_{k}^{\prime \prime }\left( t,x\right) &=&0,\ \ \ \ \ \
k=1,2,3
\label{t102} \\
\mu ^{\prime }A_{4}\left( t,x\right) +A_{1}\left( t,x\right)
-A_{4}^{\prime
}\left( t,x\right) &=&0,  \label{t103} \\
A_{0}^{\prime }\left( t,x\right) -\dot{A}_{4}\left( t,x\right) &=&0,
\label{t104} \\
\dot{A}_{0}\left( t,x\right) -A_{4}^{\prime }\left( t,x\right) &=&0.
\label{t105}
\end{eqnarray}

For $k=1$ Eq. (\ref{t101}) can be written as
\begin{equation}
\left[ e^{\mu }\dot{A}_{1}\left( t,x\right) \right] ^{\prime }=0.
\label{t106}
\end{equation}
Integrating Eq. (\ref{t106}) with respect to $x$, and then with
respect to $t$ yields
\begin{equation}
A_{1}\left( t,x\right) =f_{1}\left( t\right) e^{-\mu }+g_{1}\left(
x\right) , \label{t107}
\end{equation}
where $f_{1}\left( t\right) $ and $g_{1}\left( x\right) $ are
functions of integration.

Similarly for $k=2$, Eq. (\ref{t101})\ can be written as%
\begin{equation}
A_{2}\left( t,x\right) =f_{2}\left( t\right) e^{-\mu }+g_{2}\left(
x\right) . \label{t112}
\end{equation}
Differentiating Eq. (\ref{t107}) twice with respect to $x$ gives
\begin{equation}
A_{1}^{\prime \prime }\left( t,x\right) =\left( \mu ^{\prime 2}-\mu
^{\prime \prime }\right) e^{-\mu }f_{1}\left( t\right)
+g_{1}^{\prime \prime }\left( x\right) .
\end{equation}
Substituting the values $A_{1}^{\prime }\left( t,x\right)$ and
$A_{1}^{\prime \prime }(t,x)$ in Eq. (\ref{t98}) we get after some
simplification
\begin{equation}
\mu ^{\prime \prime }f_{1}\left( t\right) =\left[
e^{\mu}g_{1}^{\prime }\left( x\right) \right] ^{\prime } ,
\label{t108}
\end{equation}
Differentiating Eq. (\ref{t108}) with respect to $t$ we get%
\begin{equation}
\mu ^{\prime \prime }\dot{f}_{1}\left( t\right) =0 .
\end{equation}
This equation gives rise to two cases: Either $\mu ^{\prime \prime
}=0$ or not. In the first case we let $\mu =ax+b$, so that the
general form of the metric is
\begin{equation}
ds^{2}=dt^{2}-dx^{2}-e^{(ax+b)}(dy^{2}+dz^{2}) .
\end{equation}
It is a conformally flat spacetime having 15-dimensional Lie
algebra.

When $\dot{f}_{1}\left( t\right) =0$ we put $f_{1}\left( t\right)
=d_{1}$, a constant, so that Eq. (\ref{t108}) on integrating twice
with respect to $x$ gives

\begin{equation}
g_{1}\left( x\right) =d_{1}\int e^{-\mu }\mu ^{\prime }dx+d_{2}\int
e^{-\mu }dx+d_{3}.
\end{equation}
Inserting the values of $f_{1}\left( t\right) $ and $g_{1}\left(
x\right) $ in Eq.(\ref{t107}), we obtain

\begin{equation}
A_{1}\left( t,x\right) =d_{2}\int e^{-\mu }dx+d_{3}.  \label{t111}
\end{equation}%
Using this value in Eq. (\ref{t102}) gives
\begin{equation}
A_{1}\left( t,x\right) =d_{3}.  \label{t111a}
\end{equation}
Substituting the value of $A_{2}\left( t,x\right)$ from  Eq.
(\ref{t112}) in Eq. (\ref{t99}) and keeping in view that $\mu
^{\prime \prime }\neq 0$ we see that
\begin{equation}
A_{2}\left( t,x\right) =d_{4}. \nonumber
\end{equation}
Similarly,
\begin{equation}
A_{3}\left( t,x\right) =d_{5}. \nonumber
\end{equation}
Eliminating $A_{0}\left( t,x\right)$ between Eq. (\ref{t104}) and
Eq.(\ref{t105}) and using Eq. (\ref{t103}) shows that
\begin{equation}
A_{4}\left( t,x\right) =A_{1}\left( t,x\right)=0 . \nonumber
\end{equation}
Finally, Eq. (\ref{t104}) and (\ref{t105}) yield
\begin{equation}
A_{0}\left( t,x\right) =c_{1}. \nonumber
\end{equation}

Thus the conformal factor becomes zero and we get (calling $d_4$ and
$d_5$ as $c_3$ and $c_2$ respectively)
\begin{eqnarray}
\xi^{0} =c_{1}, \nonumber \\
\xi^{1} =0, \nonumber \\
\xi^{2} =c_{4}z+c_{2}, \nonumber \\
\xi^{3} =-c_{4}y+c_{3},  \nonumber 
\end{eqnarray}
which is a 4-dimensional Killing algebra representing the minimal
isometry for plane symmetric static spacetimes.

Now, we consider the case when ${\nu}^{\prime}\neq 0$ and
$\mu^{\prime }=0$. Proceeding in the same fashion as before, we see
that the constraint equations (Eqs. (\ref{t65a})-(\ref{t69})) give
rise to two possibilities: Either $(\nu ^{\prime
\prime}e^{2\nu})^{\prime }$ is zero or  not. In the first case we
write $\nu ^{\prime \prime}e^{2\nu}=k_1$, a constant, and find that
$\phi =0$, giving the KVs as

\begin{eqnarray}
\xi ^{0} &=&\left( c_{5}\cos \sqrt{k_{1}}t+c_{6}\sin
\sqrt{k_{1}}t\right)
\int e^{-2\nu}dx+c_{1}, \nonumber \\
\xi ^{1} &=&\frac{1}{\sqrt{k_{1}}}\left( c_{5}\sin
\sqrt{k_{1}}t-c_{6}\cos
\sqrt{k_{1}}t\right) , \nonumber \\
\xi ^{2} &=&c_{4}z+c_{2}, \nonumber \\
\xi ^{3} &=&-c_{4}y+c_{3} . \nonumber
\end{eqnarray}
The six dimensional Killing algebra is given by

\begin{tabular}{lll}
$\left[ \mathbf{X}_{1},\mathbf{X}_{2}\right] =0$, & $\left[
\mathbf{X}_{1}, \mathbf{X}_{3}\right] =0$, & $\left[
\mathbf{X}_{1},\mathbf{X}_{4}\right] =0$,
\\
$\left[ \mathbf{X}_{1},\mathbf{X}_{5}\right] =-k_{2}\mathbf{X}_{6}$,
& $\left[ \mathbf{X}_{1},\mathbf{X}_{6}\right]
=k_{2}\mathbf{X}_{5}$, & $\left[ \mathbf{X}_{2},\mathbf{X}_{3}\right] =0$, \\
$\left[ \mathbf{X}_{2},\mathbf{X}_{3}\right] =0$, & $\left[
\mathbf{X}_{2},\mathbf{X}_{4}\right] =-\mathbf{X}_{3}$, & $\left[
\mathbf{X}_{2},\mathbf{X}_{5}\right] =0$, \\
$\left[ \mathbf{X}_{2},\mathbf{X}_{6}\right] =0$, & $\left[
\mathbf{X}_{3},\mathbf{X}_{4}\right] =\mathbf{X}_{2}$, & $\left[
\mathbf{X}_{3},\mathbf{X}_{5}\right] =0$, \\
$\left[ \mathbf{X}_{3},\mathbf{X}_{6}\right] =0$, & $\left[
\mathbf{X}_{4}, \mathbf{X}_{5}\right] =0$, & $\left[
\mathbf{X}_{4},\mathbf{X}_{6}\right] =0$,\\
$\left[ \mathbf{X}_{5},\mathbf{X}_{6}\right] =k_{3}\mathbf{X} _{1}$.
&  &
\end{tabular}

Here $k_{2}=(x+\nu^{\prime}/\nu^{\prime \prime})$ and $k_{3}=(
\sqrt{k_{1}}\left( \int e^{-2\nu }dx\right)
^{2}+\frac{1}{\sqrt{k_{1}}}e^{-2\nu })$ are constants.

We note that when $\nu^{\prime \prime}$ is zero (i.e. $\nu^{\prime}$
is constant) we obtain 5 KVs given by
\begin{eqnarray}
\xi ^{0} &=&-\nu ^{\prime }c_{5}t+c_{1} , \nonumber \\
\xi ^{1} &=&c_{5} ,   \nonumber \\
\xi ^{2} &=&c_{4}z+c_{2} ,  \nonumber \\
\xi ^{3} &=&-c_{4}y+c_{3} . \nonumber
\end{eqnarray}
The Lie algebra in this case is

\begin{tabular}{lll}
$\left[ \mathbf{X}_{1},\mathbf{X}_{2}\right] =0$,
& $\left[ \mathbf{X}_{1},\mathbf{X}_{3}\right] =0$, & $\left[ \mathbf{X}_{1},%
\mathbf{X}_{4}\right] =0$, \\
$\left[ \mathbf{X}_{1},\mathbf{X}_{5}\right] =-\nu ^{\prime
}\mathbf{X}_{1}$, & $\left[ \mathbf{X}_{2},%
\mathbf{X}_{3}\right] =0$, & $\left[ \mathbf{X}_{2},\mathbf{X}_{4}\right] =-%
\mathbf{X}_{3}$, \\
$\left[ \mathbf{X}_{2},\mathbf{X}_{5}\right] =0$, & $\left[ \mathbf{X}_{3},%
\mathbf{X}_{4}\right] =\mathbf{X}_{2}$, & $\left[ \mathbf{X}_{3},\mathbf{X}_{5}\right] =0$, \\
$\left[ \mathbf{X}_{4},\mathbf{X}_{5}\right] =0$. &  &
\end{tabular}

In the other case $\nu^{\prime \prime}$ is not zero, and we either
get minimal symmetry for the plane or six dimensional homotheties
with $\phi =c_{4}.$

\begin{eqnarray}
\xi ^{0} &=&kc_{5}t+c_{1} ,  \nonumber \\
\xi ^{1} &=&c_{5}x,   \nonumber \\
\xi ^{2} &=&c_{4}z+c_{5}y+c_{2},  \nonumber  \\
\xi ^{3} &=&-c_{4}y+c_{5}z+c_{3}. \nonumber
\end{eqnarray}
The Lie algebra is given by

\begin{tabular}{lll}
$\left[ \mathbf{X}_{1},\mathbf{X}_{2}\right] =0$, & $\left[
\mathbf{X}_{1},
\mathbf{X}_{3}\right] =0$, & $\left[ \mathbf{X}_{1},\mathbf{X}_{4}\right] =0$, \\
$\left[ \mathbf{X}_{1},\mathbf{X}_{5}\right] =k\mathbf{X}_{1}$, &
$\left[ \mathbf{X}_{2},\mathbf{X}_{3}\right] =0$, & $\left[
\mathbf{X}_{2},\mathbf{X}_{4}\right] =-\mathbf{X}_{3}$, \\
$\left[ \mathbf{X}_{2},\mathbf{X}_{5}\right] =\mathbf{X}_{2}$, &
$\left[ \mathbf{X}_{3},\mathbf{X}_{4}\right] =\mathbf{X}_{2}$, &
$\left[ \mathbf{X}
_{3},\mathbf{X}_{5}\right] =\mathbf{X}_{3}$, \\
$\left[ \mathbf{X}_{4},\mathbf{X}_{5}\right] =0$. &  &
\end{tabular}

Here $k=-\nu ^{\prime }x+\nu ^{\prime }k_{1}+1$ is a constant.

\section{Conclusion}

Conformal motions are the vectors along which the metric tensor of a
spacetime remains invariant upto a factor, called the conformal
factor. If this factor is constant then the symmetry is called an HV
and if it is zero we get KVs. Therefore HVs and KVs are special
cases of CKVs. Here we have classified plane symmetric static
spacetimes according to conformal motion (or CKVs). We have solved
conformal Killing equations, which are first order nonhomogeneous
differential equations, to construct the general form of CKVs and
the conformal factor $\phi$ along with a set of constraint
equations. To simplify the classification scheme we divide it into
cases depending upon whether the metric coefficients $\nu(x)$ and
$\mu(x)$ are constant or not.

When both $\nu$ and $\mu$ are constant, we get flat spacetime
admitting a maximal of 15 CKVs with the conformal factor given by
\begin{equation}
\phi \left( t,x,y,z\right) =c_{5}t+c_{1}x-c_{3}y+c_{2}z+c_{6}.
\nonumber
\end{equation}

In case when $\nu ^{\prime}=0,$ but $\mu ^{\prime}\neq 0$, we get a
4-dimensional minimal Killing algebra for static plane symmetry or
the metric becomes conformally flat. The metric in this case and the
one when both $\nu ^{\prime }$ and $\mu ^{\prime }$ are nonzero, are
conformally related and give the same conformal algebra. However,
when $\nu ^{\prime }=\mu ^{\prime }\neq0$, we get a flat spacetime
with 15-dimensional algebra.

When we take $\nu ^{\prime}\neq 0,$ and $\mu ^{\prime}=0$, we obtain
5- or 6-dimensional Killing algebras or 6-dimensional homothety
algebra.

Thus we conclude that plane symmetric static spacetimes do not admit
non-trivial conformal motions apart from KVs or HVs, or the
conformally flat cases.

\acknowledgments

A research grant from the Higher Education Commission of Pakistan is
gratefully acknowledged.


\begin{thebibliography}{999}


\bibitem{2r} H. Stephani, D. Kramer, M. A. H. MacCallum, C. Hoenselaers and E.
Herlt, \textit{Exact Solutions of Einstein's Field Equations}
(Cambridge University Press, 2003).

\bibitem{9r} G. S. Hall, \textit{Symmetries and Curvature Structure in General
Relativity} (World Scientific, 2004).

\bibitem{5r} Y. Choquet-Bruhat, C. Dewitt-Morrette and M. Dillard-Bleick,
\textit{Analysis, Manifolds and Physics} (North-Holland, 1977).

\bibitem{6r} R. Maartens and S. D. Maharaj, \emph{Class. Quantum Grav.} \textbf{3} (1986) 1005.

\bibitem{7r} R. Maartens and S. D. Maharaj, \emph{Class. Quantum Grav.} \textbf{8} (1991) 503.

\bibitem{11r} R. Maartens, S. D. Maharaj and B. O. J. Tupper,
\emph{Class. Quantum Grav.} \textbf{12} (1995) 2577.

\bibitem{8r} B. O. J. Tupper, A. J. Keane, G. S. Hall, A. A. Coley and
J. Carot, \emph{Class. Quantum Grav.} \textbf{20} (2003) 801.

\bibitem{3r} A. Qadir and M. Ziad, Proceedings of the VI Marcel Grossmann
Meeting, Eds. T. Nakamura and H. Sato (World Scientific Singapore,
1993), p. 1115.

\bibitem{16r} T. Feroze, A. Qadir and M. Ziad, \emph{J. Math. Phys.}
\textbf{42} (2001) 49471.

\bibitem{thesis} Shair-e-Yazdan, \textit{Classification of Conformal Motions in
Plane Symmetric Static Spacetimes}, M.Phil. Thesis, Quaid-i-Azam
University, Islamabad (2005).

\end{thebibliography}
\end{document}